# RADIAL DISTRIBUTION OF DOSE WITHIN HEAVY CHARGED PARTICLE TRACKS – MODELS AND EXPERIMENTAL VERIFICATION USING LiF:Mg,Cu,P TL DETECTORS

W. Gieszczyk[a], P. Bilski[a], P. Olko[a] and B. Obryk[a]

[a] Institute of Nuclear Physics Polish Academy of Sciences, Radzikowskiego 152, 31-342 Krakow, Poland

## ABSTRACT

A new method of experimental verification of radial dose distribution models using solid state thermoluminescent (TL) detectors LiF:Mg,Cu,P has been recently proposed. In this work the method was applied to verify the spatial distribution of energy deposition within a single $^{131}$Xe ion track. Detectors were irradiated at the Department of Physics of the University of Jyväskylä, Finland. The obtained results have been compared with theoretical data, calculated according to the Zhang *et al.*, Cucinotta *et al.* and Geiss *et al.* radial dose distribution (RDD) models. At the lowest dose range the Zhang *et al.* RDD model exhibited the best agreement as compared to experimental data. In the intermediate dose range, up to $10^4$ Gy, the best agreement was found for the RDD model of Cucinotta *et al*. The probability of occurrence of doses higher than $10^4$ Gy within a single $^{131}$Xe ion track was found to be lower than predicted by all the studied RDD models. This may be a result of diffusion of the charge, which is then captured by TL-related trapping sites, at the distances up to dozens of nanometers from the ionization site.

## 1. INTRODUCTION

Penetrating radiation, such as gamma rays, transfers its energy to the medium indirectly. Because of its low ionization density and the mechanisms for interaction with matter this type of radiation creates a nearly uniform (over a macroscopic scale) dose distribution within a whole detector volume. In turn, the heavy charged particle (HCP) passing through matter undergoes interactions with the medium depositing energy through atomic excitations and ionizations. In this latter case secondary charged particles are produced, mainly electrons, which mediate the transfer of energy of the primary particle. The HCP creates a track of energy deposition events, producing a radial dose distribution (RDD) extending away from the track center, or core.

The RDD is one of the fundamental concepts that are used to describe tracks of heavy charged particles. Until now, the models based on this concept (Katz, 1978; Scholtz and Kraft, 1996) were mainly used to predict biological effects and the response of physical detectors after irradiation with heavy ions. Most of the analytical formulations of RDD models (Cucinotta *et al.*, 1997; Geiss *et al.*, 1998; Katz, 1978; Waligórski *et al.*, 1986) show that at radial distances lower than about 1 nm from the path of a heavy ion track the local dose can reach values as high as $10^6$ Gy. This values have never been verified experimentally and thus especially no one has ever tried to use TL detectors for this purpose, because of too low a spatial resolution as well as the fact that the response of most TL materials saturates typically around the dose of $10^3$ Gy.

As it was demonstrated before (Olko *et al.*, 2011), about 20% of the entire energy of a 5.5 MeV alpha particle, penetrating LiF, is deposited within local doses exceeding 50 kGy. It was also showed that, by analysis of the LiF:Mg,Cu,P (MCP-N) TL detector glow curves, up to 550 $^{o}$C, it is possible to identify the local doses within a single alpha particle track as high as MGy. Based on this feature a new method of experimental verification of RDD models using solid state MCP-N TL detectors has been recently proposed (Gieszczyk *et al.*, 2012). The method has already been tested for alpha particles and the RDD model developed by Geiss *et al.* (1998). In this work the method was applied to verify the spatial distribution of energy deposition within a single $^{131}$Xe ion track. Verification was performed based on the experimental data, obtained from the readout of highly irradiated MCP-N TL detectors, and theoretical data calculated according to the RDD models developed by Zhang *et al.* (1985), Cucinotta *et al.* (1997) and Geiss *et al.* (1998) (see Table 1 and Fig. 2A).

## 2. MATERIALS AND METHODS

*2.1. The approach*

The MCP-N TL glow curves measured after irradiations with either gamma rays or heavy charged particles can be deconvoluted into identical sets of individual peaks differing only in their amplitudes (Gieszczyk *et al.*, 2012), as it is shown in Fig. 1. It is also assumed that these amplitudes are related as follows:

$$A_i^{ion} \pm \delta A_i^{ion} = C \cdot \int_{D_{min}}^{D_{max}} A_i^\gamma(D) \cdot f(D) \cdot dD \qquad (1)$$

where: $A_i^{ion}$ – the amplitude of the *i*-peak measured for heavy charged particles with uncertainty $\delta A_i^{ion}$, $A_i^\gamma(D)$ – dose-response dependent peak amplitude measured for the same *i*-peak after irradiation with gamma rays, $f(D)$ – unknown function describing frequency of occurrence of a given dose $D$ within a single particle track, $C$ – proportionality factor, $i = 1 \ldots k$, $k$ – number of individual peaks. The radial dose distribution is defined as a point dose $D$ at the radial distance $r$ from the track axis. It is assumed here that ion paths are straight lines. $D_{max}$ is defined for $r \to 0$, while $D_{min}$ corresponds to the dose deposited at the maximum range of $\delta$-rays, $r_\delta$. Having measured $A_i^{ion}$, $\delta A_i^{ion}$ and $A_i^\gamma(D)$ the $f(D)$ function can be unfolded and compared with theoretical data, calculated according to the analyzed RDD models. Within this work unfolding was performed using iterative SAND II algorithm (McElroy *et al.*, 1967).

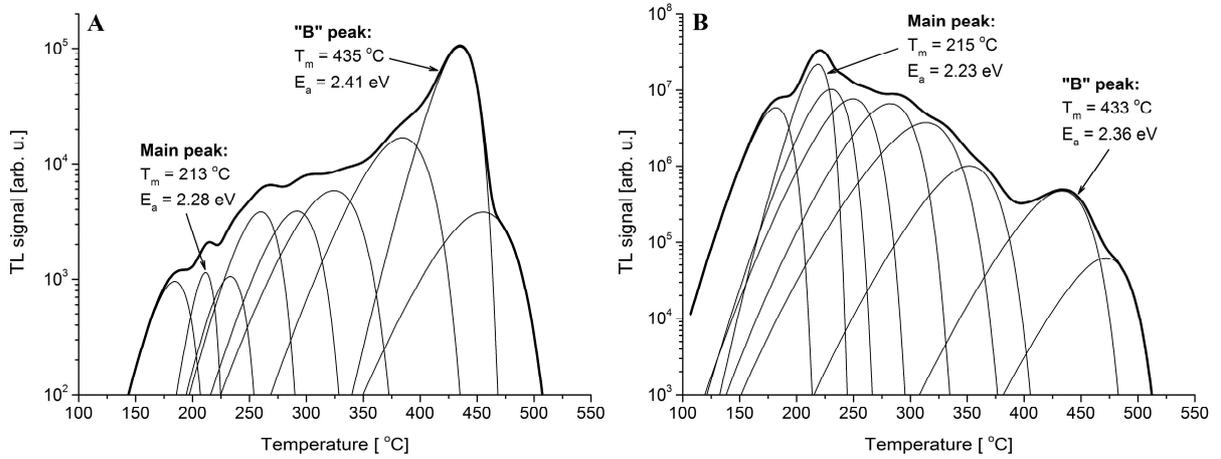

Fig. 1. MCP-N TL glow curves deconvolution. Glow curves measured for gamma rays dose of 284 kGy (panel A) and for the $^{131}$Xe ions fluence of $2 \cdot 10^9$ cm$^{-2}$ (panel B). High local doses, within a single charged particle track, cause the appearance of high-temperature peaks characteristic for doses exceeding 100 kGy of gamma radiation.

*2.2. Calculation of the initial response functions*

The iterative unfolding code requires the use of the initial response function. This function is iteratively modified until minimum fitting error (equation (5)) is reached. Because the algorithm is sensitive to initial parameters the free-form initial response function cannot be used. Therefore, starting functions for the analyzed RDD models were calculated numerically using a track-segment approach. The track has been divided into $n = 1000$ segments in the axial direction. It is assumed that within each segment the particle energy $E_j$ (j = 1, …, n) remains constant. The segment thickness, $h_j$, was defined as the difference between the range of particle of the energy $E_j$ and $E_{j+1}$. Ion ranges for particular energies were calculated using SRIM code (Ziegler *et al.*, 2010). All segments had the same thickness on a logarithmic scale. Each segment was also divided into $m = 1000$ dose ranges (from $D_l$ to $D_{l+1}$, $l = 0 \ldots m$) in the radial direction. Within each *j* track segment the probability of occurrence of doses from a given range, $f(D_l, D_{l+1})$, was calculated according to the following formula:

$$f(D_l, D_{l+1}) = \frac{1}{V_t} \cdot \sum_{j=1}^{n} \frac{\pi \left[ r(D_l, E_j)^2 - r(D_{l+1}, E_j)^2 \right] \cdot h_j}{d_l} \tag{2}$$

where: $r(D_l, E_j)$ – distance from the track core, in which the dose $D_l$ occurs, within the track segment of energy $E_j$, and $d_l = D_{l+1}-D_l$. $r(D_l, E_j)$ values were calculated according to the analyzed RDD models using the *libamtrack* library (Greilich *et al.*, 2010). $V_t$ denotes the volume of the track, which was calculated as follows:

$$V_t = \sum_{j=1}^{n} \pi \cdot r_\delta(E_j)^2 \cdot h_j \tag{3}$$

where: $r_\delta(E_j)$ – maximum range of delta electrons emitted by the primary particle of the energy $E_j$, within the $j$ track segment, calculated according to a given electron range model (see Table 1). The initial response functions, calculated for the analyzed RDD models according to equations (2) and (3), are compared in Fig. 2B. It is visible that the Cucinotta *et al.* model predicts a lower probability of occurrence of the lowest doses within a single particle track, as compared to the other analyzed models. In turn, according to this RDD model doses higher than $10^2$ Gy occur within a single particle track with higher probability regarding the other analyzed models. Data presented in Fig. 2B have been used as the initial conditions for the iterative unfolding procedure. It should be mentioned that conversion of the unfolded response function into the RDD function is not trivial, because knowing the probability of occurrence of the dose $D$ within a whole Xe ion track, determination of the probability of occurrence of this dose within each a track segment is ambiguous. Therefore, in this work the probability of occurrence of the dose $D$ within a single particle track, calculated according to the studied RDD models, will be compared to the corresponding response functions unfolded using experimental data, according to equation (1).

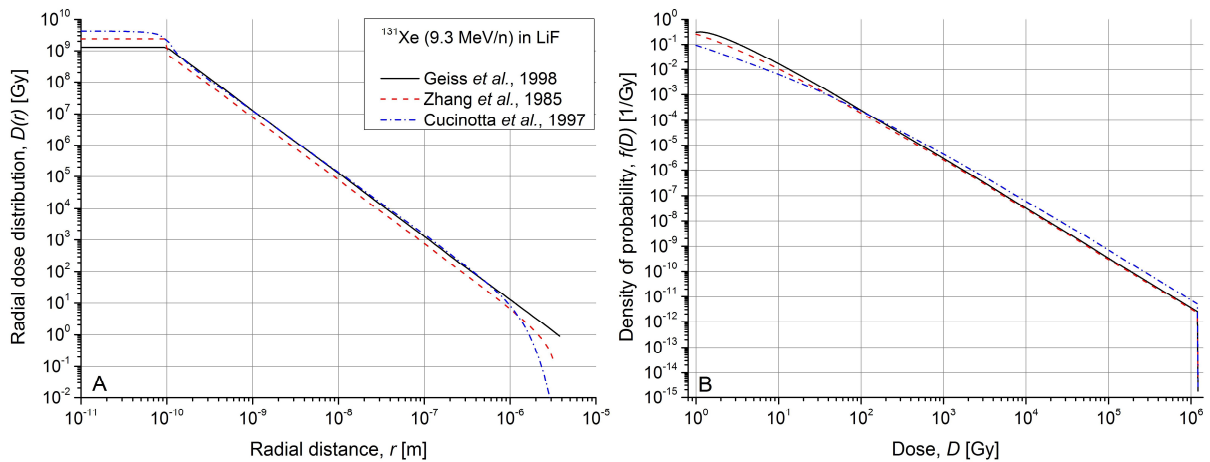

Fig. 2. Panel A: analyzed radial dose distribution models. Panel B: probability of occurrence of a given dose $D$ within a single particle track (initial response functions calculated for the analyzed RDD models according to equation (2)). Data obtained by the usage of *libamtrack* library (Greilich *et al.*, 2010).

Table 1. Analyzed models of radial dose distribution and energy-range relationship for delta electrons. Definition of symbols can be found in (Grzanka, 2013).

| | Electron range model | Radial dose distribution formula |
|---|---|---|
| Zhang *et al.* (1985) | $r_\delta(E) = l \cdot E^\alpha$ | $D(r) = C \frac{Z^{*2}}{\beta^2} \frac{1}{\rho} \frac{1}{r^2} \frac{1}{\alpha} \left(1 - \frac{r}{r_\delta}\right)^{\frac{1}{\alpha}}$ |
| Tabata *et al.* (1972) Cucinotta *et al.* (1997) | $r_\delta(E) = a_1 \left( \frac{1}{a_2} \ln\left(1 + a_2 \frac{E}{m_e c^2}\right) - \frac{a_3 \frac{E}{m_e c^2}}{1 + a_4 \left(\frac{E}{m_e c^2}\right)^{a_5}} \right)$ | $D(r) = C \frac{Z^{*2}}{\beta^2} \frac{f_s(r) \cdot f_l(r)}{\rho} \frac{1}{r^2} + N(LET) \frac{1}{\rho} \frac{exp\left(-\frac{r}{2d}\right)}{r^2}$ |
| Geiss *et al.* (1998) | $r_\delta(E) = \sqrt{E^3}$ | $D(r) = \begin{cases} k, r < r_0 \\ k \frac{r_0^2}{r^2}, r_0 < r \leq r_\delta \end{cases}$ |

*2.3. Experimental conditions*

Irradiations of MCP-N TL detectors were implemented at the Department of Physics of the University of Jyväskylä, Finland (Gieszczyk *et al.*, 2013a). 9.3 MeV/n xenon ions were chosen for irradiations, since we can expect that ions of higher ionization density will deposit higher local doses around the track core. Therefore, high-dose, high-temperature effects observed in the MCP-N TL glow curves should be better displayed. Detectors were exposed using a fluence range from $10^5$ to $10^9$ particles/cm$^2$, which is equivalent to the dose range from $10^0$ to $2 \cdot 10^4$ Gy (in the meaning of an average dose deposited in the irradiated detector volume, calculated according to Gieszczyk *et al.*, 2013a). The samples have been read out using the Harshaw 3500 manual TLD reader, up to 550 $^o$C, at a constant heating rate of 2 $^o$C/s. The dose-response dependence for individual TL peaks measured for gamma radiation was performed by reanalysis and deconvolution of the glow curves measured by Obryk et al. (2009) and by analysis and deconvolution of the glow curves measured by Gieszczyk *et al.* (2013b, c). All measured glow curves have been deconvoluted into individual 1$^{st}$ order TL peaks using the *GlowFit* code (Puchalska and Bilski, 2006).

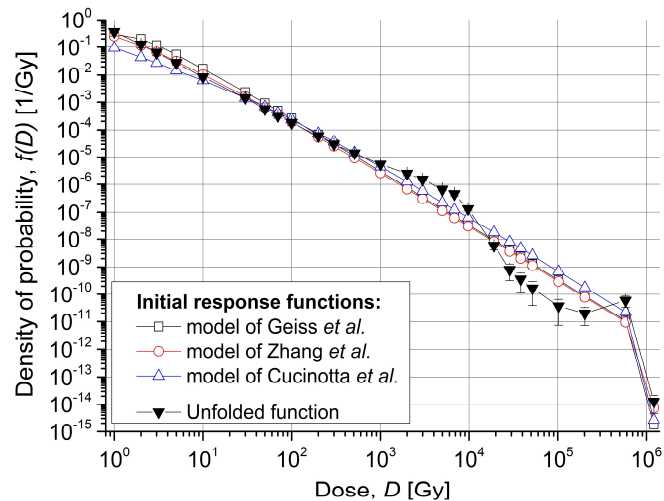

Fig. 3. Comparison of initial and unfolded response functions for the analyzed RDD models. Results indicate lower probability of occurrence of doses higher than $10^4$ Gy within a single particle track in comparison to the model data.

## 3. RESULTS AND DISCUSSION

Fig. 3 compares an arithmetically averaged unfolded response function with initial response functions calculated for the analyzed RDD models. The averaged function was applied since resulting response functions weakly depend on the initial response function, as was expected. At the lowest dose range, up to 10 Gy, the RDD model of Zhang *et al.* shows the best agreement with experimental data. Theoretical data points for the models of Geiss *et al.* and Cucinotta *et al.* seem to be over- and underestimated in this dose range, respectively. In the intermediate dose range, up to $10^3$ Gy, good agreement between experimental and model data was found for the model of Cucinotta *et al.* At the end of this dose range, Geiss *et al.* and Zhang *et al.* models predictions are underestimated. Between $10^3$ and $10^4$ Gy unfolded response function exceeds predictions of the analyzed RDD models. Obtained results tend to suggest that even if a certain fraction of the entire energy of a 9.3MeV/n $^{131}$Xe ion is deposited within local doses exceeding $10^4$ Gy, the probability of occurrence of such high doses within a single particle track is significantly lower, as compared to the analyzed RDD models predictions. Such a result, for doses exceeding $10^3$ Gy, may result from diffusion of the charge, from the areas located close to the ion path, which is then captured by TL-related trapping sites, at the distances up to dozens of nanometers from the ionization site. This was suggested by microdosimetric model of one-hit detector for LiF:Mg,Cu,P (Olko, 2002).

Table 2. Relative percentage difference between experimental data and values calculated with the initial and unfolded response functions. Newly proposed procedure gives results consistent with experimental data within 15%.

| Peak No. | Peak amplitude (experiment) $A_i^{ion}$ [arb. u.] | Relative percentage difference between experimental and calculated data [%] | | | | | |
| --- | --- | --- | --- | --- | --- | --- | --- |
| | | Zhang model (Initial function) | Zhang model (Unfolded function) | Cucinotta model (Initial function) | Cucinotta model (Unfolded function) | Geiss model (Initial function) | Geiss model (Unfolded function) |
| 2 | 3429254 | 24.66 | 1.83 | 6.79 | 15.33 | 41.94 | 1.06 |
| 3 | 1066338 | 22.49 | 7.28 | 7.11 | 2.42 | 41.62 | 12.39 |
| 4 | 31398793 | 1.72 | 12.95 | 1.82 | 2.31 | 57.06 | 18.13 |
| 5 | 3301999 | 41.33 | 3.11 | 33.63 | 14.71 | 34.61 | 2.93 |
| 6 | 2254179 | 60.10 | 6.76 | 29.19 | 3.03 | 30.18 | 0.50 |
| 7 | 593738 | 81.22 | 7.01 | 43.80 | 1.51 | 6.88 | 0.48 |
| 8 | 172490 | 150.67 | 4.26 | 39.21 | 0.32 | 33.31 | 3.06 |
| 9 | 25758 | 24.81 | 5.77 | 59.90 | 5.42 | 75.69 | 8.85 |
| 10 | 4435 | 88.93 | 3.00 | 3.93 | 5.68 | 11.59 | 13.19 |
| 11 | 1004 | 122.33 | 5.25 | 19.47 | 1.10 | 25.22 | 0.60 |
| Quality index, $Q$ | | 5.92 | 0.04 | 0.95 | 0.05 | 1.65 | 0.07 |

In the Table 2 relative percentage differences, $\Delta_i$ [%]:

$$\Delta_i\ [\%] = \left(\frac{A_i^{exp} - A_i^{calc}}{A_i^{calc}}\right) \cdot 100\% \qquad (4)$$

between experimental and calculated data are presented. The calculated values were obtained by solving equation (1), when initial response functions as well as unfolded response functions were utilized. One can see that using initial response functions discrepancies between experimental and calculated data are very high. Replacing this function by the unfolded response function the differences lower than 10% have been achieved. Only in a few cases the relative percentage error has reached higher values. The quality index, $Q$:

$$Q = \sum_{i=1}^{N} \left[ \frac{A_i^{exp} - A_i^{calc}}{A_i^{calc}} \right]^2 \qquad (5)$$

which was minimized during the iterative unfolding procedure, has been reduced below the value of 0.1 (see Table 2) for all analyzed RDD models. In general, one may conclude that newly proposed method gives results consistent with experimental data within 15%.

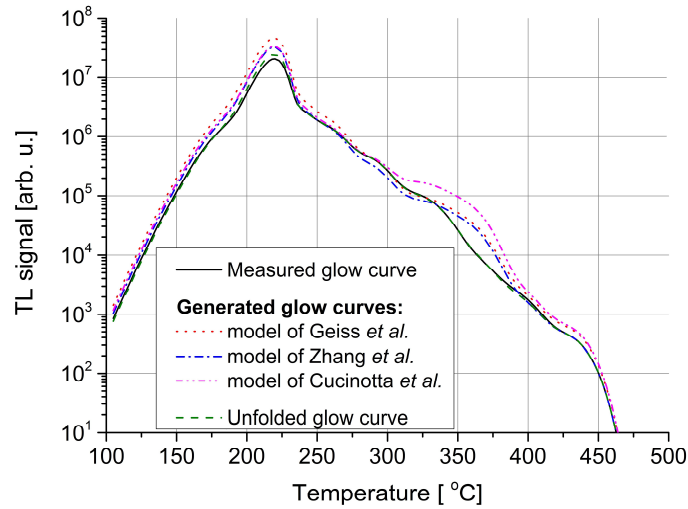

Fig. 4. Comparison of the experimentally measured TL glow curve and the glow curves artificially generated using single TL peaks amplitudes, calculated by solving equation (1) for both initial and unfolded response functions.

By solving equation (1), for both initial and unfolded response functions, the amplitudes of individual TL peaks have been obtained. Assuming that other parameters (activation energy and x-axis position) describing a single TL peak remain constant, the shape of corresponding TL glow curves can be simulated. In Fig. 4 the artificially generated TL glow curves, for three analyzed RDD models, are compared to the glow curve experimentally measured after exposure of the studied detectors to $^{131}$Xe ions. It is clearly visible that the usage of the unfolded response function generates a TL glow curve, which is almost identical to the measured one. This result confirms that the method is suitable for verification of RDD models predictions.

## 4. CONCLUDING REMARKS

The new method of experimental verification of radial dose distribution models using solid state TL detectors LiF:Mg,Cu,P has been proposed. Within this work the method was tested for 9.3 MeV/n $^{131}$Xe ion beam and RDD models developed by Zhang *et al.* (1985), Cucinotta *et al.* (1997) and Geiss *et al.* (1998). The results indicate that probability of occurrence of doses below 10 Gy, within a single particle track, is well described by the RDD model of Zhang *et al.* The RDD model of Cucinotta *et al.* presents good agreement with experimental data at the intermediate dose range. Even if a certain part of the entire energy of a 9.3 MeV/n Xe ion is deposited within local doses exceeding $10^4$ Gy, the probability of occurrence of such high doses within a single particle track is significantly lower in comparison to the analyzed RDD models predictions. Such a dependence may result from diffusion of the charge, which is then captured by TL-related trapping sites, at the distances up to dozens of nanometers from the ionization site, as was predicted by microdosimetric model of one-hit detector for LiF:Mg,Cu,P (Olko, 2002). The newly proposed method gives results consistent with experimental data within 15%.

## ACKNOWLEDGEMENTS

Wojciech Gieszczyk has been partly supported by the EU Human Capital Operation Program, Polish Project No. POKL. 04.0101-00-434/08-00.

## REFERENCES

Cucinotta, F., Wilson, J.W., Shavers, M.R. and Katz, R., 1997. Calculation of heavy ion inactivation and mutation rates in radial dose model of track structure. NASA Technical Paper 3630.

Geiss, O.B., Kramer, M., Kraft, G., 1998. Efficiency of thermoluminescent detectors to heavy charged particles. Nucl. Instr. And Meth. in Phys. Res. B 142, 592–598.

Gieszczyk, W., Olko, P., Bilski, P., Grzanka, L., Obryk, B., Horwacik, T., 2012. Application of LiF:Mg,Cu,P (MCP-N) thermoluminescent detectors (TLD) for experimental verification of radial dose distribution models. Nukleonika 57 (4), 507–512.

Gieszczyk, W., Bilski, P., Olko, P., Herrmann, R., Kettunen, H., Virtanen, A., Bassler, N., 2013a. Evaluation of the relative thermoluminescence efficiency of LiF:Mg,Ti and LiF:Mg,Cu,P TL detectors to low-energy heavy ions. Radiat. Meas. 51-52, 7-12.

Gieszczyk, W., Bilski, P., Obryk, B., Olko, P. and Bos, A. J. J., 2013b. Measurements of high-temperature emission spectra of highly irradiated LiF:Mg,Cu,P (MCP-N) TL detectors. Radiat. Meas. 56, 183-186.

Gieszczyk, W., Bilski, P., Obryk, B., Olko, P. and Bos, A. J. J., 2013c. Spectral characteristic of high-dose high-temperature emission from LiF:Mg,Cu,P (MCP-N) TL detectors. Radiat. Meas. 53-54, 22-30.

Greilich, S., Grzanka, L., Bassler, N., Andersen, C. E., Jäkel, O., 2010. Amorphous track models: A numerical comparison study. Radiat. Meas. 45, 1406-1409.

Grzanka, L., 2013. Modelling beam transport and biological effectiveness to develop treatment planning for ion beam radiotherapy. Ph.D Thesis, IFJ PAN, Krakow. http://www.ifj.edu.pl/msd/rozprawy_dr/rozpr_Grzanka.pdf

Katz, R., 1978. Track structure theory in radiobiology and in radiation protection. Nucl. Track Detection 2, 1–28.

McElroy WN, Berg S, Crockett T, Hawkins RG, 1967. A computer automated iterative method for neutron flux spectra determination by foil activation. AFWL-TR-67-41.

Obryk, B., Bilski, P., Budzanowski, M., Fuerstner, M., Glaser, M., Ilgner, C., Olko, P., Pajor, A., Stuglik, Z., 2009. Development of a method for passive measurement of radiation doses at ultra-high dose range. IEEE Trans. Nucl. Sci. 56, 3759-3763.

Olko, P., 2002. Microdosimetric modelling of physical and radiobiological detectors. Habilitation Thesis, IFJ PAN, Krakow, Report No 1914/D. http://www.ifj.edu.pl/publ/reports/2002/1914.pdf?lang=pl

Olko, P., Bilski, P., Gieszczyk, W., Grzanka, L., Obryk, B., 2011. Microdosimetric analysis of response of LiF:Mg,Cu,P (MCP-N) TL detectors for alpha-particles and ultra-high doses of gamma-rays. Radiat. Meas. 46, 1349–1353.

Puchalska, M., Bilski, P., 2006. GlowFit – a new tool for thermoluminescence glow curve deconvolution, Radiat. Meas. 41, 659-664.

Scholz, M., Kraft, G., 1996. Track structure and the calculation of biological effects of heavy charged particles. Adv. Space Res. 18, 5e14.

Tabata, T., Ito, R., Okabe, S., 1972. Generalized semiempirical equations for the extrapolated range of electrons. Nucl. Instrum. And Methods 103, 91-95.

Waligórski, M.P.R., Hamm, R.N., Katz, R., 1986. The radial distribution of dose around the path of a heavy ion in water. Nucl. Tracks Radiat. Meas. 11, 309–319.

Zhang, C., Dunn, E., Katz, R., 1985. Radial distribution of dose and cross section for inactivation of dry enzymes and viruses. Radiat. Protect. Dosim. 13, 215-218.

Ziegler, J.F., Ziegler, M.D., Biersack, J.P., 2010. SRIM – the stopping and range of ions in matter. Nucl. Instr. Methods Phys. Res. B 268, 1818-1823.